\documentclass[
twocolumn,
]{ceurart}

\usepackage{listings}
\usepackage[utf8]{inputenc} 
\usepackage{subcaption}
\usepackage{adjustbox,lipsum}
\usepackage{amsmath}

\begin{document}

\copyrightyear{2022}
\copyrightclause{Copyright for this paper by its authors.
  Use permitted under Creative Commons License Attribution 4.0
  International (CC BY 4.0).}

\conference{EvalRS 2022: CIKM EvalRS 2022 DataChallenge, October 21, 2022, Atlanta, GA 
}

\title{Item-based Variational Auto-encoder for Fair Music Recommendation}
\author[1]{Jinhyeok Park}[%
email=jinhyeok1234@postech.ac.kr,
]\fnmark[1]

\author[1]{Dain Kim}[%
email=dain5832@postech.ac.kr,
]\fnmark[1]

\author[1]{Dongwoo Kim}[%
email=dongwookim@postech.ac.kr,
]\cormark[1]

\cortext[1]{Corresponding author.}
\fntext[1]{These authors contributed equally.}
\address[1]{Pohang University of Science and Technology, Pohang, Republic of Korea}

\begin{abstract}
We present our solution for the EvalRS DataChallenge. 
The EvalRS DataChallenge aims to build a more realistic recommender system considering accuracy, fairness, and diversity in evaluation.
Our proposed system is based on an ensemble between an item-based variational auto-encoder (VAE) and a Bayesian personalized ranking matrix factorization (BPRMF).
To mitigate the bias in popularity, we use an item-based VAE for each popularity group with an additional fairness regularization. 
To make a reasonable recommendation even the predictions are inaccurate, we combine the recommended list of BPRMF and that of item-based VAE.
Through the experiments, we demonstrate that the item-based VAE with fairness regularization significantly reduces popularity bias compared to the user-based VAE. The ensemble between the item-based VAE and BPRMF makes the top-1 item similar to the ground truth even the predictions are inaccurate.
Finally, we propose a `Coefficient Variance based Fairness' as a novel evaluation metric based on our reflections from the extensive experiments. 
\end{abstract}


\maketitle

\section{Introduction}\label{introduction}
Recommender systems are rising as a powerful tool that predicts user preferences based on past interactions between users and items. Industries such as e-commerce, music, and social media adopt recommender systems to provide users with a more personalized experience and foster a marketplace. However, several works have shown that excessive emphasis on user utility alone may result in problems like the Matthew effect and filter bubble~\cite{chen2020bias, wang2022survey, morik2020controlling}. 

Utility-focused model selection is undesirable since it may lead to inequality in distribution, which eventually suppresses market diversity~\cite{singh2018fairness}. For this reason, many previous studies have proposed the necessity of metrics beyond accuracy, such as fairness, diversity, and serendipity~\cite{yao2017beyond, vargas2011rank, ge2010beyond}. For instance, \citet{li2021user} demonstrate that there exists a performance gap between the inactive and active user groups and suggest the definition of user-oriented group fairness. \citet{biega2018equity} propose equity of attention that requires the exposure to be proportional to the relevance of an item.

EvalRS DataChallenge is designed to emphasize the importance of measuring recommendation performance from various perspectives, including accuracy, fairness, and diversity~\cite{tagliabue2022evalrs}.
Using the LFM-1b dataset~\cite{schedl2016lfm}, participants are asked to recommend top-$k$ items for each user. 
The performances of recommendations are evaluated through accuracy metrics (e.g., hit rate, mean reciprocal rank), accuracy metrics on a per-group basis to measure fairness, and behavioral tests to measure the diversity of recommended items using Reclist~\cite{chia2022beyond}.
The challenge is divided into two phases depending on how each metric is aggregated. In phase 1, the final evaluation score is computed using a simple average, and in phase 2, the weight of each metric is adjusted according to the difficulty observed during phase 1 before aggregation. 
 
In this work, we propose a framework that can satisfy various evaluation metrics comprehensively. 
We adopt the variational auto-encoder for collaborative filtering~\cite{liang2018variational} as our baseline, which aims to produce a likelihood of the user-item interaction matrix from multinomial distribution via an auto-encoding architecture. 
Through extensive model evaluation, we found three strategies that can mitigate potential biases while keeping a relatively high utility.
First, we found that the item-based VAE helps to alleviate the popularity bias of recommendations compared to the user-based VAE.  
Second, we found that training separate VAE models for artist popularity groups can mitigate the popularity bias.
Lastly, we found that a fairness regularizer, designed to minimize the gap between the losses of different groups, further leverages the fairness in item groups. 

The rest of the paper is structured as follows. In \autoref{sec:method}, we describe our model architecture and strategies to improve model fairness. In \autoref{sec:experiments}, we show the experimental result with a discussion.
In \autoref{sec:discussion}, we reflect on our findings and propose a new metric that can better measure the fairness of the model by taking accuracy into account.

\paragraph{Evaluation metrics}
We describe the evaluation metrics used in the EvalRS DataChallenge~\cite{tagliabue2022evalrs}. 
The evaluation metrics can be categorized into three different measures:
\begin{itemize}
    \item \textbf{Accuracy metrics}: accuracy metrics indicate the predictive performance of a model. It includes hit rate (HR) and mean reciprocal rank (MRR), which are widely used in recommender systems.
    \item \textbf{Accuracy metrics on a per-group basis}: group-based metrics are designed to evaluate the fairness and robustness of the model. The challenge adopts the miss rate equality difference (MRED), which measures the average difference between the miss rate (MR) of each group and the MR of the entire dataset. The metrics are evaluated across five different groups: gender, country, user history, artist popularity, and track popularity.
    \item \textbf{Behavioral tests}: behavioral tests measure the similarity between recommended and ground truth items and the diversity of recommended items. Behavioral tests consist of two metrics; `be less wrong' and `latent diversity.' Be less wrong measures the distance between the embeddings of ground truth and the predicted result. Latent diversity indicates a model density in the latent space of tracks.
\end{itemize}

\section{Method}\label{method}
\label{sec:method}
\subsection{Baseline Models}
We use the variational auto-encoders (VAE) and Bayesian personalized ranking matrix factorization (BPRMF) as our backbone methods. In this section, we describe the backbone methods and explain how to use these backbones to curate the final recommendation list.

\paragraph{Variational auto-encoders for collaborative filtering.}

In this work, we employ the variational auto-encoder (VAE) for collaborative filtering~\cite{liang2018variational} as the first backbone model. The objective of VAE~\cite{kingma2013auto} is to maximize the evidence lower bound (ELBO) for each data point $x_i$:
\begin{align*}\label{eq:elbo}
L_\beta(x_i; \theta, \phi) = & \mathbb{E}_{q_\phi(z_i\mid x_i)}[\log p_\theta(x_i\mid z_i)] \\
&-\beta\cdot \mathrm{KL}(q_\phi(z_i\mid x_i) \| p(z_i)),
\end{align*}
where $z_i$ is the latent variable, $\beta$ measures the importance of the KL divergence, and the likelihood function $p_\theta$ and the variational distribution $q_\phi$ are parameterized by $\theta$ and $\phi$, respectively.

There have been multiple approaches to employ the VAE framework for collaborative filtering~\cite{liang2018variational,sedhain2015autorec}. In this work, we follow the framework proposed by \cite{liang2018variational}.

Let $\mathbf{x}_u=[x_{u1}, x_{u2}, ... x_{uI}]$ be an implicit feedback of user $u$ where $x_{ui}$ is binary indicator specifying whether user $u$ interacted with the item $i$. The likelihood function $p(\mathbf{x}_u|z_u)$ is then modeled via a multinomial distribution conditioned on the latent vector $\mathbf{z}_u$. The multivariate normal distribution is used as a variational distribution $q(\mathbf{z}_u|\mathbf{x}_u)$. During the training, one can optimize the parameters to maximize the ELBO. After training, the recommended items are chosen based on the multinomial distribution among the items that have not been interacted so far.

\paragraph{Item-based VAE} 
Although using implicit feedback of a user, i.e., $\mathbf{x}_u$, as an input of VAE is a common approach (user-based VAE), alternatively, one can use implicit feedback of an item as an input of VAE (item-based VAE). The implicit feedback vector of item $i$ can be constructed as $\mathbf{x}_i=[x_{i1}, x_{i2}, ... x_{iU}]$, where $x_{ij}$ indicates the interaction between item $i$ and user $j$.

To recommend items with item-based VAE, the model infers the interaction probability over all items to complete the user-item interaction matrix and recommends top-$N$ items for each user. Empirically, we find that the item-based VAE tends to recommend unpopular items compared to the user-based VAE.

\paragraph{Bayesian personalized ranking matrix factorization}
We use Bayesian personalized ranking matrix factorization (BPRMF)~\cite{rendle2012bpr} as the second baseline model. BPRMF estimates the posterior distribution over the likelihood of pair-wise ranking between items with a prior distribution. 

\subsection{Model Optimization}
\label{sec:optimization}
In this section, we introduce various methods used to improve the performance of the item-based VAE for phase 2. Our approach mainly targets group-based metrics and behavioral tests rather than accuracy metrics. 

\paragraph{Popularity-aware training based on items}
We aim to improve the MRED between track popularity groups and artist popularity groups, which are significant factors in phase 2. 

Based on the item-based VAE, to reduce the performance gap between artist popularity groups, we divide items by artist popularity groups and train a VAE for each group separately. After training, we find that the least popular artist group is underfitted compared to other groups. Therefore, we train two more epochs for this group. Then, we pick a certain number of items from each group to make a recommendation. Please check the details of this process in the Final Recommendation part.

The MRED between track popularity groups is also an important factor for phase 2. Although we divided items by artist groups, MRED between the track population groups is also reduced. However, the least popular item group whose playcount is less than ten is still not recommended well, as shown in \autoref{fig:1}. To recommend an item from the least popular track group, we additionally train a separate VAE for this group and include at least one item from this group.

\begin{table*}[t]
\resizebox{\textwidth}{!}{
\begin{tabular}{crrrrrrrrrr}
\toprule
                & \makecell{Hit\\Rate} & MRR & \makecell{Country\\(MRED)} & \makecell{User\\(MRED)}& \makecell{TrackPop\\(MRED)} & \makecell{ArtistPop\\(MRED)} & \makecell{Gender\\(MRED)} & \makecell{Be less\\Wrong} & \makecell{Latent\\Diversity} & \makecell{Score\\Phase1} \\ 
\midrule
VAE(item)     &\textbf{0.2121}&\textbf{0.0399}&-0.0248& -0.0287&-0.0529&-0.0216&-0.0144&0.3189&-0.3041&\textbf{0.0138}      \\ 
VAE(user)     &0.1593&0.0256&-0.0161&-0.0323&-0.0937&-0.0430&\textbf{-0.0044}&0.3512&\textbf{-0.2726}&0.0082\\ 
BPRMF          & 0.0372 & 0.0025 & \textbf{-0.0098} & \textbf{-0.0163} &\textbf{-0.0230}&\textbf{-0.0102}&-0.0070&\textbf{0.3721} &-0.2948 & 0.0056       \\
\bottomrule
\end{tabular}
}
  \caption{Phase 1 results of our baseline models obtained by simple averaging of nine metrics.}
  \label{tab:1}
\vspace*{-3mm}
\end{table*}

\begin{table}[t]
\begin{adjustbox}{width=1\columnwidth}
\begin{tabular}{crrrrrr}
\toprule
             Model & 1 & 10 & 100 & 1000 & total  \\ 
\midrule
VAE (item)     &0.8946&0.7865&0.7770& 0.8803&0.7879&\\ 
VAE (user)     &0.9398&0.8861&0.8062& 0.6448&0.8407&\\ 
BPRMF         &0.9965&0.9830&0.9387& 0.9487&0.9628&\\
\bottomrule
\end{tabular}
\end{adjustbox}
  \caption{MR for each model at each track popularity group.}
  \label{tab:2}
\vspace*{-3mm}
\end{table}

\paragraph{Fairness Regularization}
Fairness regularization aims to introduce an additional regularizer term to the objective to narrow the gap between group losses. The approach has been widely adopted in fields such as computer vision, natural language processing, and sound~\cite{zhao2019regularface, zemel2013learning}. 

Many recommender systems also employ regularizers to improve group fairness~\cite{yao2017beyond, beutel2019fairness}. 

We incorporate a fairness regularization into VAE based on the work of~\cite{borges2022f2vae}. The regularization term computes the average difference between the group reconstruction loss and the entire reconstruction loss as

\begin{equation*}\label{eq:fair_reg}
\begin{split}
F_\phi(x_i; \theta, \phi)=\mathbb{E}\left[\left|\frac{1}{\left|G_j\right|} \sum_{c \in G_j} \mathbb{E}\left[\log p_\theta\left(x_c \mid z_c\right)\right] \right.\right. \\
\left.\left. - \frac{1}{|I|} \sum_{i \in I} \mathbb{E}\left[\log p_\theta\left(x_i \mid z_i\right)\right]\right|\right],
\end{split}
\end{equation*}
where $I$ is a set of all items, and $G_j$ is a set of items that belong to the group $j$. Groups are divided into 1, 10, 100, and 1000 based on track popularity, and each item is assigned to the group according to its total play counts. Our final objective can be expressed as follows:
\begin{equation*}\label{eq:fair_loss}
L^R_\beta(x_i; \theta, \phi) = L_\beta(x_i; \theta, \phi) - \gamma\cdot F_\phi,
\end{equation*}
where the hyperparameter $\gamma$ controls the weight of the regularizer. The higher value of $\gamma$ indicates that the model takes a greater proportion of fairness into account during the optimization process.

\paragraph {Final Recommendation}

Since there are four artist popularity groups, we train four separate VAEs each of which is designated for each group. 
From the four VAEs, we first create a list of 98 items to be recommended. First, we take 38/20/20/20 items from artists groups 0, 1, 2, and 3 respectively, where group 0 indicates the least popular group. Among those selected items, we take the five most probable items from each group and curate a list of the top 20 items. The 20 items are ordered with 2, 1, 3, 0 (5/5/5/5) in descending order of the number of items in each group. The remaining 78 items are listed after with the same order (15/15/15/33). One additional item recommended from the least popular track group is added at the end of the list.

In addition, we find that the item-based VAE often fails to achieve good performance in the behavioral tests, especially for `be less wrong'. Therefore, we ensemble the item-based VAE and BPRMF which shows good performance in `be less wrong'. Since the metric only considers the top-1 item, we put the most probable item from the BPRMF model at the top of our previous recommendation list, resulting in 100 recommended items.

\section{Experiments}\label{experiments}
\label{sec:experiments}

\begin{table*}[t]
\resizebox{\textwidth}{!}{
\begin{tabular}{crrrrrrrrrrr}
\toprule
                & \makecell{Hit\\Rate}& MRR & \makecell{Country\\(MRED)} & \makecell{User\\(MRED)}& \makecell{TrackPop\\(MRED)} & \makecell{ArtistPop\\(MRED)} & \makecell{Gender\\(MRED)} & \makecell{Be less\\Wrong} & \makecell{Latent\\Diversity} &  \makecell{Score\\Phase2} \\
\midrule
Fold1     &0.0154&0.0015&-0.0030&-0.0035&-0.0021&-0.0007&-0.0003&0.3661&-0.2924\\ 
Fold2     &0.0151&0.0016&-0.0036&-0.0021&-0.0024&-0.0021&-0.0012&0.3602&-0.3000\\ 
Fold3     &0.0169&0.0021&-0.0047&-0.0044&-0.0023&-0.0005&-0.0004&0.3685&-0.2948\\   
Fold4     &0.0169&0.0017&-0.0036&-0.0017&-0.                                        0024&-0.0010&-0.0008&0.3609&-0.2984\\ \hline \hline
Average     &0.0161&0.0017&-0.0037&-0.0029&-0.0023&-0.0010&-0.0007&0.3639&-0.2964&1.553\\ 
Baseline  &0.0363&0.0037&-0.0090&-0.0224&-0.0111&-0.0072&-0.0061&0.3758&-0.3080&-1.212\\

\bottomrule
\end{tabular}
}
  \caption{Our final results for the four folds and average of them. Baseline denotes `CBOWRecSysBaseline' provided by the challenge organizers.}
  \label{tab:3}
\vspace*{-3mm}
\end{table*}

\subsection{Dataset}
The LFM-1b dataset~\cite{schedl2016lfm} is provided for the challenge. The dataset consists of the listening history of users with demographic information, such as gender and nationality, and metadata of the item. The dataset includes 119,555 users, 820,998 tracks and 37,926,429 interactions. The test set was generated based on the leave-one-out framework by randomly masking one item from each user's history. Please check the detailed pre-processing steps of the dataset for the challenge in \cite{tagliabue2022evalrs}.

\subsection{Phase 1}
\label{sec:preliminary}
In phase 1, we conduct an experiment to check the performances of our baseline models: item-based VAE, user-based VAE, and BPRMF.
For all experiments with VAEs, we adopt the same architecture as \cite{liang2018variational}. We set the batch size to 32. Latent dimension is set to 500, and the size of hidden layer is set to 300. We train for 5 epochs using the Adam ~\cite{kingma2014adam} optimizer with a learning rate of 0.001. Dropout rate is set to 0.2.
For BPRMF, we set the batch size to 8192, and dimension to 64. We train for 10 epochs with a learning rate of 0.001. Instead of using weight regularization, we normalize the vector if the maximum value of user vectors and item vectors is greater than 1.

\autoref{tab:1} reveals variational auto-encoders for collaborative filtering achieve good performance when applying a simple average of all metrics. In particular, the item-based VAE shows good performance in not only hit rate and MRR but also MRED between user activity, track popularity, and artist popularity groups. \autoref{tab:2} shows the MR of each track popularity group between baseline models. We can observe that the item-based VAE has better accuracy with unpopular item groups. 

\begin{figure}[t]
\begin{center}
\includegraphics[scale=0.25]{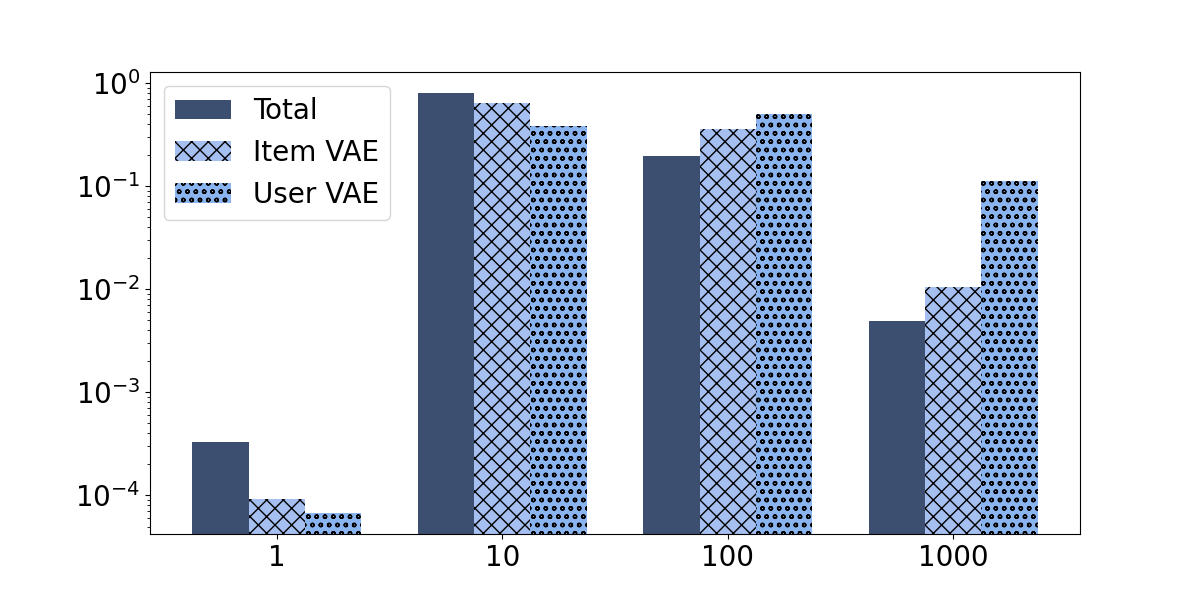}
\caption{Popularity distributions of all items and items recommended through the model. The x-axis represents track popularity groups, and y-axis represents proportion of each group. We use the logarithmic scale on the y-axis due to skewed distributions. The result shows that the recommended items from the item-VAE follow a similar distribution to the total popularity distributions of all items.}
\vspace*{-6mm}
\label{fig:1}
\end{center}
\end{figure}

We observe that the item-based VAE recommends as many unpopular items as popular ones. \autoref{fig:1} shows popularity distributions of the recommended items and all items. The result indicates that the recommended items from item-based VAE follow a similar distribution of the total item popularity computed in the training set. In the meanwhile, the user-based VAE tends to recommend more popular items than the item-based VAE.

We find that although the BPRMF method does not show a good overall performance and it tends to recommend famous items, it outperforms other methods in `Be less wrong'. \autoref{tab:1} shows the preliminary results.

\subsection{Phase 2}
Based on the preliminary results shown in phase 1, we combine the results of the item-based VAE and the BPRMF to curate the recommendation list for phase 2.

In phase 1, the overall score is determined by a simple average of each test. However, in phase 2, the importance of each metric is adjusted based on the performance of participants in phase 1. As we replicate the process described in the challenge to analyze relative values of weights, we observe an enormous gap between the weight of artist popularity and that of HR. Thus, we mainly focused on mitigating the bias of `artist popularity'.

For the final experiments, we set the latent dimension of the item-based VAE to 17 and the batch size to 32. Then, we train the model using the Adam optimizer with a learning rate of 1e-3 for two epochs. As the unpopular artist group does not fit well, We find that having additional two epochs to train the VAE for the unpopular artist group generally helps to improve the performance. When we train the least popular item group, we set the latent dimension to 15 and train for 2 epochs.
$\beta$ and $\gamma$, which are the coefficient of the KL divergence term and the coefficient of the regularizer, are set to 0.0001 and 0.003, respectively, after parameter searching. For BPRMF, we set the latent dimension to 200 and other conditions are set to the same as the baseline. Then, we make the final recommendation described in \ref{sec:optimization}. \footnote{The source code for reproducing the experiments is available at https://github.com/ParkJinHyeock/evalRS-submission}

\autoref{tab:3} shows our results of all folds with the phase 2 score and the baseline score. The results show that our model successfully reduces the gap between the artist popularity groups and the track popularity groups. Furthermore, our model shows lower MREDs between user activity, gender, and country groups than those of the baseline provided by the challenge organizers.

\begin{table*}[t]
\begin{tabular}{c|c|rrrrrrrr}
\toprule
 & Model & 1 & 10 & 100 & 1000 & Hit & MRED  & CV(ours)\\ 
\midrule
\multirow{4}{*}{\makecell{Track\\popularity}} &VAE (item)&0.8946&0.7865&0.7770& 0.8803&0.2121&-0.0529&0.2559\\ 
&                 VAE (user)     &0.9398&0.8861&0.8063& 0.6448&0.1593&-0.0937&0.7022\\ 
&                 VAE (final)    &0.9858&0.9851&0.9821& 0.9867&0.0161&-0.0023&0.0701\\ 
&                 BPRMF           &0.9965&0.9831&0.9387& 0.9487&0.0372&-0.0230&0.6436\\  
\bottomrule
\end{tabular}

\begin{tabular}{c|c|rrrrrrr}
\toprule
             &Model & 1 & 100 & 1000 & 10000 & Hit  & MRED   & CV(ours)\\ 
\midrule
\multirow{4}{*}{\makecell{Artist\\popularity}}&VAE (item)&0.8259&0.8107&0.7688&0.7942&0.2121&-0.0216&0.1019\\ 
&VAE (user)     &0.8962&0.8887&0.8556&0.7870&0.1593&-0.0430&0.2716\\ 
&VAE (final)    &0.9831&0.9848&0.9835&0.9841&0.0161&-0.0010&0.1459\\ 
&BPRMF          &0.9850&0.9721&0.9629&0.9546&0.0372&-0.0102&0.3070\\ 

\bottomrule
\end{tabular}
  \caption{Miss rate and proposed metric of each model at each track popularity group (top) and artist popularity group (bottom). VAE (final) denotes our final submission.}
  \label{tab:4}
\vspace*{-3mm}
\end{table*}

\section{Discussion and Reflection}\label{discussion}
\label{sec:discussion}
\subsection{User Fairness}
As shown in experiments, our main approach focuses on balancing the HR of the artist and track popularity groups. However, our method also yields fair performance in user-related fairness metrics. The previous study~\cite{ekstrand2018all} shows that there are no significant differences in the performance of the model between the gender groups. The authors also identify a negative relationship between user activity and performance. We observe a similar phenomenon from our results; there are small differences between gender and country groups, and a negative correlation between user activity and performance. However, with the item-based VAE, reducing the gap between item groups also reduces the gap between user activity groups.
\subsection{Reflection on Evaluation Metric} %
The EvalRS DataChallenge~\cite{tagliabue2022evalrs} evaluates the fairness of the model using the average difference of MR between groups. In this section, we analyze the weakness of this approach and propose a novel metric that improves it.

We first analyze the limitation of the current fairness metric.
Suppose there is a model with a hit rate of 0.2 and another with a hit rate of 0.02. If the average deviation of HR is both 0.01, the two models would be considered to produce equivalent performance regarding fairness. However, in terms of relative ratios, the same deviation accounts for 5\% of the former but 50\% of the latter. From this perspective, using MRED to measure fairness might lead to unreasonable comparison.

With this intuition, we propose a `Coefficient of Variance (CV)\cite{brown1998coefficient} based fairness' which is less sensitive to scale. The Coefficient of Variance is defined as the standard deviation divided by the mean multiplied by 100:

\begin{equation} \label{eq:2} 
    \text{CV} = \frac{\sigma}{m} * 100 
\end{equation}

By dividing by the mean, the metric indicates the relative ratio of deviation to performance. Inspired by this, our proposed metric can be expressed as follows.

\begin{equation} \label{eq:3}
    \text{CV}_{\text{HR}} = (\text{HR}_{\text{avg}})^{-1} \sqrt{\cfrac{\sum_{i}(\text{HR}_{\text{avg}} - \text{HR}_{\text{group}_{i}})^{2}}{N_{\text{groups}}}}
\end{equation}

The proposed metric quantifies the fairness of the model, considering the average of HR when measuring the deviation. The lower value indicates higher fairness. Our metric reasonably evaluates fairness through a relative ratio. Even if the model achieves a low deviation, the proposed metric will be penalized if the deviation is relatively large compared to the HR. \autoref{tab:4} shows the MR of each group, MRED, and our proposed metric. We observe that for artist popularity groups, the item-based VAE outperforms the final model as it has a relatively low deviation with high HR, which is consistent with our intuition. Models with a relatively large deviation between each group have a high penalty, while models with a relatively low deviation have a low penalty.

\section{Conclusion}\label{conclusion}
In this work, we propose a fairness-aware variational auto-encoder for recommender systems. Our approach shows that the item-based VAE significantly reduces the popularity bias of the model. Moreover, we conclude that obtaining the recommendation results from various artist groups and adapting a regularizer further improves the fairness of the model. Finally, we suggest the notion of `Coefficient of Variance based Fairness' for the model evaluation and demonstrate that it reasonably measures the fairness of the model.

\bibliography{ref}

\begin{thebibliography}{23}
\expandafter\ifx\csname natexlab\endcsname\relax\def\natexlab#1{#1}\fi
\providecommand{\url}[1]{\texttt{#1}}
\providecommand{\href}[2]{#2}
\providecommand{\path}[1]{#1}
\providecommand{\DOIprefix}{doi:}
\providecommand{\ArXivprefix}{arXiv:}
\providecommand{\URLprefix}{URL: }
\providecommand{\Pubmedprefix}{pmid:}
\providecommand{\doi}[1]{\href{http://dx.doi.org/#1}{\path{#1}}}
\providecommand{\Pubmed}[1]{\href{pmid:#1}{\path{#1}}}
\providecommand{\bibinfo}[2]{#2}
\ifx\xfnm\relax \def\xfnm[#1]{\unskip,\space#1}\fi
\bibitem[{Chen et~al.(2020)Chen, Dong, Wang, Feng, Wang, and He}]{chen2020bias}
\bibinfo{author}{J.~Chen}, \bibinfo{author}{H.~Dong},
  \bibinfo{author}{X.~Wang}, \bibinfo{author}{F.~Feng},
  \bibinfo{author}{M.~Wang}, \bibinfo{author}{X.~He},
\newblock \bibinfo{title}{Bias and debias in recommender system: A survey and
  future directions},
\newblock \bibinfo{journal}{arXiv preprint arXiv:2010.03240}
  (\bibinfo{year}{2020}).
\bibitem[{Wang et~al.(2022)Wang, Ma, Zhang*, Liu, and Ma}]{wang2022survey}
\bibinfo{author}{Y.~Wang}, \bibinfo{author}{W.~Ma},
  \bibinfo{author}{M.~Zhang*}, \bibinfo{author}{Y.~Liu},
  \bibinfo{author}{S.~Ma},
\newblock \bibinfo{title}{A survey on the fairness of recommender systems},
\newblock \bibinfo{journal}{ACM Journal of the ACM (JACM)}
  (\bibinfo{year}{2022}).
\bibitem[{Morik et~al.(2020)Morik, Singh, Hong, and
  Joachims}]{morik2020controlling}
\bibinfo{author}{M.~Morik}, \bibinfo{author}{A.~Singh},
  \bibinfo{author}{J.~Hong}, \bibinfo{author}{T.~Joachims},
\newblock \bibinfo{title}{Controlling fairness and bias in dynamic
  learning-to-rank},
\newblock in: \bibinfo{booktitle}{Proceedings of the 43rd international ACM
  SIGIR conference on research and development in information retrieval},
  \bibinfo{year}{2020}, pp. \bibinfo{pages}{429--438}.
\bibitem[{Singh and Joachims(2018)}]{singh2018fairness}
\bibinfo{author}{A.~Singh}, \bibinfo{author}{T.~Joachims},
\newblock \bibinfo{title}{Fairness of exposure in rankings},
\newblock in: \bibinfo{booktitle}{Proceedings of the 24th ACM SIGKDD
  International Conference on Knowledge Discovery \& Data Mining},
  \bibinfo{year}{2018}, pp. \bibinfo{pages}{2219--2228}.
\bibitem[{Yao and Huang(2017)}]{yao2017beyond}
\bibinfo{author}{S.~Yao}, \bibinfo{author}{B.~Huang},
\newblock \bibinfo{title}{Beyond parity: Fairness objectives for collaborative
  filtering},
\newblock \bibinfo{journal}{Advances in neural information processing systems}
  \bibinfo{volume}{30} (\bibinfo{year}{2017}).
\bibitem[{Vargas and Castells(2011)}]{vargas2011rank}
\bibinfo{author}{S.~Vargas}, \bibinfo{author}{P.~Castells},
\newblock \bibinfo{title}{Rank and relevance in novelty and diversity metrics
  for recommender systems},
\newblock in: \bibinfo{booktitle}{Proceedings of the fifth ACM conference on
  Recommender systems}, \bibinfo{year}{2011}, pp. \bibinfo{pages}{109--116}.
\bibitem[{Ge et~al.(2010)Ge, Delgado-Battenfeld, and Jannach}]{ge2010beyond}
\bibinfo{author}{M.~Ge}, \bibinfo{author}{C.~Delgado-Battenfeld},
  \bibinfo{author}{D.~Jannach},
\newblock \bibinfo{title}{Beyond accuracy: evaluating recommender systems by
  coverage and serendipity},
\newblock in: \bibinfo{booktitle}{Proceedings of the fourth ACM conference on
  Recommender systems}, \bibinfo{year}{2010}, pp. \bibinfo{pages}{257--260}.
\bibitem[{Li et~al.(2021)Li, Chen, Fu, Ge, and Zhang}]{li2021user}
\bibinfo{author}{Y.~Li}, \bibinfo{author}{H.~Chen}, \bibinfo{author}{Z.~Fu},
  \bibinfo{author}{Y.~Ge}, \bibinfo{author}{Y.~Zhang},
\newblock \bibinfo{title}{User-oriented fairness in recommendation},
\newblock in: \bibinfo{booktitle}{Proceedings of the Web Conference 2021},
  \bibinfo{year}{2021}, pp. \bibinfo{pages}{624--632}.
\bibitem[{Biega et~al.(2018)Biega, Gummadi, and Weikum}]{biega2018equity}
\bibinfo{author}{A.~J. Biega}, \bibinfo{author}{K.~P. Gummadi},
  \bibinfo{author}{G.~Weikum},
\newblock \bibinfo{title}{Equity of attention: Amortizing individual fairness
  in rankings},
\newblock in: \bibinfo{booktitle}{The 41st international acm sigir conference
  on research \& development in information retrieval}, \bibinfo{year}{2018},
  pp. \bibinfo{pages}{405--414}.
\bibitem[{Tagliabue et~al.(2022)Tagliabue, Bianchi, Schnabel, Attanasio, Greco,
  Moreira, and Chia}]{tagliabue2022evalrs}
\bibinfo{author}{J.~Tagliabue}, \bibinfo{author}{F.~Bianchi},
  \bibinfo{author}{T.~Schnabel}, \bibinfo{author}{G.~Attanasio},
  \bibinfo{author}{C.~Greco}, \bibinfo{author}{G.~d. S.~P. Moreira},
  \bibinfo{author}{P.~J. Chia},
\newblock \bibinfo{title}{Evalrs: a rounded evaluation of recommender systems},
\newblock \bibinfo{journal}{arXiv preprint arXiv:2207.05772}
  (\bibinfo{year}{2022}).
\bibitem[{Schedl(2016)}]{schedl2016lfm}
\bibinfo{author}{M.~Schedl},
\newblock \bibinfo{title}{The lfm-1b dataset for music retrieval and
  recommendation},
\newblock in: \bibinfo{booktitle}{Proceedings of the 2016 ACM on international
  conference on multimedia retrieval}, \bibinfo{year}{2016}, pp.
  \bibinfo{pages}{103--110}.
\bibitem[{Chia et~al.(2022)Chia, Tagliabue, Bianchi, He, and
  Ko}]{chia2022beyond}
\bibinfo{author}{P.~J. Chia}, \bibinfo{author}{J.~Tagliabue},
  \bibinfo{author}{F.~Bianchi}, \bibinfo{author}{C.~He},
  \bibinfo{author}{B.~Ko},
\newblock \bibinfo{title}{Beyond ndcg: behavioral testing of recommender
  systems with reclist},
\newblock in: \bibinfo{booktitle}{Companion Proceedings of the Web Conference
  2022}, \bibinfo{year}{2022}, pp. \bibinfo{pages}{99--104}.
\bibitem[{Liang et~al.(2018)Liang, Krishnan, Hoffman, and
  Jebara}]{liang2018variational}
\bibinfo{author}{D.~Liang}, \bibinfo{author}{R.~G. Krishnan},
  \bibinfo{author}{M.~D. Hoffman}, \bibinfo{author}{T.~Jebara},
\newblock \bibinfo{title}{Variational autoencoders for collaborative
  filtering},
\newblock in: \bibinfo{booktitle}{Proceedings of the 2018 world wide web
  conference}, \bibinfo{year}{2018}, pp. \bibinfo{pages}{689--698}.
\bibitem[{Kingma and Welling(2013)}]{kingma2013auto}
\bibinfo{author}{D.~P. Kingma}, \bibinfo{author}{M.~Welling},
\newblock \bibinfo{title}{Auto-encoding variational bayes},
\newblock \bibinfo{journal}{arXiv preprint arXiv:1312.6114}
  (\bibinfo{year}{2013}).
\bibitem[{Sedhain et~al.(2015)Sedhain, Menon, Sanner, and
  Xie}]{sedhain2015autorec}
\bibinfo{author}{S.~Sedhain}, \bibinfo{author}{A.~K. Menon},
  \bibinfo{author}{S.~Sanner}, \bibinfo{author}{L.~Xie},
\newblock \bibinfo{title}{Autorec: Autoencoders meet collaborative filtering},
\newblock in: \bibinfo{booktitle}{Proceedings of the 24th international
  conference on World Wide Web}, \bibinfo{year}{2015}, pp.
  \bibinfo{pages}{111--112}.
\bibitem[{Rendle et~al.(2012)Rendle, Freudenthaler, Gantner, and
  Schmidt-Thieme}]{rendle2012bpr}
\bibinfo{author}{S.~Rendle}, \bibinfo{author}{C.~Freudenthaler},
  \bibinfo{author}{Z.~Gantner}, \bibinfo{author}{L.~Schmidt-Thieme},
\newblock \bibinfo{title}{Bpr: Bayesian personalized ranking from implicit
  feedback},
\newblock \bibinfo{journal}{arXiv preprint arXiv:1205.2618}
  (\bibinfo{year}{2012}).
\bibitem[{Zhao et~al.(2019)Zhao, Xu, and Cheng}]{zhao2019regularface}
\bibinfo{author}{K.~Zhao}, \bibinfo{author}{J.~Xu}, \bibinfo{author}{M.-M.
  Cheng},
\newblock \bibinfo{title}{Regularface: Deep face recognition via exclusive
  regularization},
\newblock in: \bibinfo{booktitle}{Proceedings of the IEEE/CVF Conference on
  Computer Vision and Pattern Recognition}, \bibinfo{year}{2019}, pp.
  \bibinfo{pages}{1136--1144}.
\bibitem[{Zemel et~al.(2013)Zemel, Wu, Swersky, Pitassi, and
  Dwork}]{zemel2013learning}
\bibinfo{author}{R.~Zemel}, \bibinfo{author}{Y.~Wu},
  \bibinfo{author}{K.~Swersky}, \bibinfo{author}{T.~Pitassi},
  \bibinfo{author}{C.~Dwork},
\newblock \bibinfo{title}{Learning fair representations},
\newblock in: \bibinfo{booktitle}{International conference on machine
  learning}, \bibinfo{organization}{PMLR}, \bibinfo{year}{2013}, pp.
  \bibinfo{pages}{325--333}.
\bibitem[{Beutel et~al.(2019)Beutel, Chen, Doshi, Qian, Wei, Wu, Heldt, Zhao,
  Hong, Chi et~al.}]{beutel2019fairness}
\bibinfo{author}{A.~Beutel}, \bibinfo{author}{J.~Chen},
  \bibinfo{author}{T.~Doshi}, \bibinfo{author}{H.~Qian},
  \bibinfo{author}{L.~Wei}, \bibinfo{author}{Y.~Wu},
  \bibinfo{author}{L.~Heldt}, \bibinfo{author}{Z.~Zhao},
  \bibinfo{author}{L.~Hong}, \bibinfo{author}{E.~H. Chi}, et~al.,
\newblock \bibinfo{title}{Fairness in recommendation ranking through pairwise
  comparisons},
\newblock in: \bibinfo{booktitle}{Proceedings of the 25th ACM SIGKDD
  International Conference on Knowledge Discovery \& Data Mining},
  \bibinfo{year}{2019}, pp. \bibinfo{pages}{2212--2220}.
\bibitem[{Borges and Stefanidis(2022)}]{borges2022f2vae}
\bibinfo{author}{R.~Borges}, \bibinfo{author}{K.~Stefanidis},
\newblock \bibinfo{title}{F2vae: a framework for mitigating user unfairness in
  recommendation systems},
\newblock in: \bibinfo{booktitle}{Proceedings of the 37th ACM/SIGAPP Symposium
  on Applied Computing}, \bibinfo{year}{2022}, pp. \bibinfo{pages}{1391--1398}.
\bibitem[{Kingma and Ba(2014)}]{kingma2014adam}
\bibinfo{author}{D.~P. Kingma}, \bibinfo{author}{J.~Ba},
\newblock \bibinfo{title}{Adam: A method for stochastic optimization},
\newblock \bibinfo{journal}{arXiv preprint arXiv:1412.6980}
  (\bibinfo{year}{2014}).
\bibitem[{Ekstrand et~al.(2018)Ekstrand, Tian, Azpiazu, Ekstrand, Anuyah,
  McNeill, and Pera}]{ekstrand2018all}
\bibinfo{author}{M.~D. Ekstrand}, \bibinfo{author}{M.~Tian},
  \bibinfo{author}{I.~M. Azpiazu}, \bibinfo{author}{J.~D. Ekstrand},
  \bibinfo{author}{O.~Anuyah}, \bibinfo{author}{D.~McNeill},
  \bibinfo{author}{M.~S. Pera},
\newblock \bibinfo{title}{All the cool kids, how do they fit in?: Popularity
  and demographic biases in recommender evaluation and effectiveness},
\newblock in: \bibinfo{booktitle}{Conference on fairness, accountability and
  transparency}, \bibinfo{organization}{PMLR}, \bibinfo{year}{2018}, pp.
  \bibinfo{pages}{172--186}.
\bibitem[{Brown(1998)}]{brown1998coefficient}
\bibinfo{author}{C.~E. Brown},
\newblock \bibinfo{title}{Coefficient of variation},
\newblock in: \bibinfo{booktitle}{Applied multivariate statistics in
  geohydrology and related sciences}, \bibinfo{publisher}{Springer},
  \bibinfo{year}{1998}, pp. \bibinfo{pages}{155--157}.

\end{thebibliography}
\end{document}